\documentclass[a4paper,11pt]{amsart}
\usepackage{amsmath, amsthm}
\usepackage{amscd}
\usepackage{amssymb}
\usepackage{array}
\usepackage{color}
\usepackage{enumerate}
\usepackage{enumitem}
\usepackage{MnSymbol}
\usepackage{bbm}
\usepackage{amsfonts}

\newcommand{\salg}{\mathrm{(salg)}}
\newcommand{\sets}{\mathrm{(sets)}}
\usepackage{graphics}
\newcommand{\cX}{\mathcal{X}}

\theoremstyle{definition}

\newcommand{\bM}{{\bf M}}
\newcommand{\eb}{{\bold e}}
\newcommand{\beq}{\begin{equation}}
\newcommand{\ee}{\end{equation}}

\newcommand{\1}{\mathbbm{1}}

\newcommand{\R}{\mathbb{R}}
\newcommand{\C}{\mathbb{C}}

\newcommand{\bP}{\mathbb{P}}

\newcommand{\cL}{\mathcal{L}}

\newcommand{\ep}{\mathcal{E}}

\newcommand{\rSpO}{\widetilde{\mathrm{SpO}}}

\newcommand{\tSp}{\widetilde{\mathrm{Sp}}}
\newcommand{\tsp}{\widetilde{\mathrm{sp}}}
\newcommand{\rGL}{\mathrm{GL}}

\newcommand{\rSU}{\mathrm{SU}}
\newcommand{\rSL}{\mathrm{SL}}

\newcommand{\rsl}{\mathrm{sl}}

\newcommand{\al}{\alpha}
\newcommand{\be}{\beta}

\newcommand{\lra}{\longrightarrow}

\begin{document}

\title{The symplectic origin of conformal and Minkowski superspaces}

\maketitle
\centerline{ R. Fioresi$^\dagger$, E. Latini$^\star$}

\medskip
\centerline{\it $^\dagger$ Dipartimento di Matematica, Universit\`{a} di
Bologna }
 \centerline{\it Piazza di Porta S. Donato, 5. 40126 Bologna. Italy.}
\centerline{\footnotesize e-mail: rita.fioresi@UniBo.it}

\medskip
\centerline{\it $^\star$ Institut f{\"u}r Mathematik}
 \centerline{\it Universit{\"a}t Z{\"u}rich-Irchel,}
  \centerline{\it Winterthurerstrasse 190, CH-8057 Z{\"u}rich, Switzerland}
\centerline{{\footnotesize e-mail: 
emanuele.latini@math.uzh.ch}}

\vspace{10pt}

\vspace{10pt}

\renewcommand{\arraystretch}{1}


\begin{abstract} 
Supermanifolds provide a very natural ground to understand and handle supersymmetry from a geometric point of view; supersymmetry in $d=3,4,6$ and $10$ dimensions is also deeply related to the normed division algebras.

In this paper we want to show the link between the conformal group and certain types of symplectic transformations over division algebras. Inspired by this observation we then propose a new\,realization of the real form of the 4 dimensional conformal and Minkowski superspaces we obtain, respectively, as a Lagrangian supermanifold over the twistor superspace $\mathbb{C}^{4|1}$ and a big cell inside it. 
 The beauty of this approach is that it naturally generalizes to the 6 dimensional case (and possibly also to the 10 dimensional one) thus providing an elegant and uniform characterization of the conformal superspaces.

\vspace{2cm}

\end{abstract}

\pagestyle{myheadings} \markboth{Fioresi, Latini}{Symplectic Minkowski superspaces and division algebras}

\newpage

\tableofcontents

\section{Introduction}\label{intro}
Supersymmetry (SUSY) is a part of the modern approach to the theory of elementary particles;  supersymmetric string theory offers in fact the most promising model, so far, for the unification of all the elementary forces in a manner compatible with quantum theory and general relativity.\\
SUSY can be naturally treated by packing all physical fields, taking values in the 4d Minkowsli space ${\bf M}^{3,1}$, into a unique object, \emph{i.e.} the superfield, that is assumed to take values in the 4d Minkowski superspace ${\bf M}^{3,1|1}$, on which one has locally commuting and anticommuting coordinates.

For mathematicians supersymmetry, the concept of superspace and supermanifolds were inspiring and gave a new look at geometry, both differential and algebraic.  In geometry, in fact, "objects" are built out of local pieces: the most general of such object is a superspace, and the symmetries of such an object are then supersymmetries which are described by supergroups. The  
functor of points originally introduced by Grothendieck to study algebraic
geometry, is now an essential tool to recover the geometric nature of 
supergroups and superspaces, which is otherwise difficult to grasp
through the sheaf theoretic approach. It is actually and surprisingly the
point of view physicists took at the very beginning of this theory,
when points of a superspace were understood with the use of grassmann
algebras, which are nothing but superalgebras over a superspace consisting
of a point (see \cite{be}). 
This suggestive point of view was later fully explained and
justified by Shvarts and Voronov in \cite{sh,vo} and then linked to
the theory of functor of points a la Grothendieck in \cite{bcf1, bcf2}.
Later Manin in \cite{ma1,ma2} took full advantage of the machinery of
the functor of points and applied it to the theory of superspaces
and superschemes in particular to develop the theory of superflags
and supergrassmannians, which is of particular interest to us.
In the present work however, we shall make an effort to leave the full machinery
of functor of points on the background, though employing its power in
the description of the $T$-points of a supergroup or a superspace.
Hence we shall rely for the results on the works \cite{ccf} and
\cite{fl}, where all of the foundations of the theory are fully explained
together with their physical significance.

Another relevant invariance principle in physics is given by the conformal symmetry; many physical systems, as those for massless particles, enjoy this symmetry and one may then imagine that there are regimes where conformal invariance is restored. In conformal geometry physics is described by equivalence classes of metrics so that all equations are manifestly locally Weyl invariant.

 Minkowski space time, by the way, is not
enough to support conformal symmetry, and it needs to be compactified by
adding points at infinity to obtain a space endowed with a natural action of the conformal
group. This is evident from the Dirac cone construction in which one considers the  space of light like rays in ${\bf M}^{d,2}$ (known as the conformal space or conformal sphere) and the compactified Minkowski space $\overline{{\bf M}}^{d-1,1}$ is then realized as one particular section of the cone.
The above flat model for conformal geometry was generalized in \cite{FG} by Fefferman and Graham  who replaced ${\bf M}^{d,2}$ by a $d+2$ dimensional manifold equipped with a metric which admits a hypersurface orthogonal homothety. Note that there is a natural interpretation of this picture as curved Cartan geometry; moreover the Cartan approach naturally leads to a Weyl covariant differential calculus, known as tractor calculus, originally constructed in \cite{tractor} (see also \cite{Gover:2008sw,Curry:2014yoa} for a physics oriented review) and then generalized to all parabolic geometries in \cite{tractorparabolic}. It can be understood as the equivalent of the superfield formalism for conformal invariance.

In this paper we aim to point out that the conformal space in $3,4,6$ or 10 dimensions may be also understood as a certain Lagrangian manifold over the four normed division algebras $\mathbb{K}=\mathbb{R},\mathbb{C},\mathbb{H}, \textrm{and}\,\mathbb{O}$; we will use the notation $n:=\textrm{dim}\,\mathbb{K}=1,2,4 \,\textrm{and}\, 8$ and denote by $\mathbb{K'}$ the split version of any division algebra. The relationship between SUSY, sipertwistors and $\mathbb{K}$  is also a recurring theme \cite{Baez:2009xt,cederwall1,cederwall2,cederwall3}. For example nonabelian Yang-Mills theories are supersymmetric only if the dimension of the Minkowski spacetime is $d= 3, 4, 6$ or 10 (and the same is true for the Green-Schwarz superstring). In this case SUSY relies on the vanishing of a certain trilinear expression that it turns to be strictly related to the existence of the four normed division algebras in $d-2$ dimensions \cite{evans}. Recently in \cite{Anastasiou:2013cya} Duff and collaborators used normed division algebras to give a descr
 iption of supergravity by "tensoring" super Yang-Mills multiplets\footnote{In literature there are in fact many attempts to understand the quantum theory of gravity through the idea of "Gravity is the square of Yang-Mills theories" idea supported by the fact that (super)gravity scattering amplitudes can be obtained from those of (super)Yang-Mills \cite{Bern:2010ue}.}. The main argument the authors used  is the observation that the entries of second row of the $2\times 2$ half split magic square \cite{MS,MS2,BS}
\vspace{1mm}
\begin{center}
\begin{tabular}{ccccc}
\hline\\[-4mm]
&$\mathbb{R}$&$\mathbb{C}$&$\mathbb{H}$&$\mathbb{O}$\\[.1mm]\hline\\[-3mm]
$\mathbb{C}'$&$\mathfrak{so}(2,1)$&$\mathfrak{so}(3,1)$&$\mathfrak{so}(5,1)$&$\mathfrak{so}(9,1)$
\end{tabular}
\end{center}
\vspace{1mm}
can be naturally represented as $\mathfrak{sl}_2(\mathbb{K})$ producing then the Lie algebras isomorphisms
$$
\mathfrak{sl}_2({\mathbb{K}})=\mathfrak{so}(n+1,1)\,.$$

We believe that understanding the relation between SUSY and normed division algebras from a supergeometric point of view could also give a fundamental contribution to the study of the quantum properties of supergravity.  

In a series of papers\,\cite{fl,flv}  the complex 4 dimensional Minkowski (super)space was realized as the big cell inside a complex flag (super)manifold where the conformal group $\mathrm{SL}_4(\mathbb{C})$ acts naturally
while the real Minkowski (super)space was then obtained as a suitable real form of the complex version. We observe that this approach essentially coincides with the Dirac cone construction; we plan then to take advantages of this observation in a future project where we will try to study supergeometry within the Cartan approach.  \\
In \cite{Kuzenko:2012tb}, the authors constructed the compactified Minkowsky 3d superspace $\overline{{\bf M}}^{2,1}$ and its supersymmetric extension, as a Lagrangian manifold over
the twistor space $\mathbb{R}^4$; this relies to the isomorphism $\mathrm{Spin}(3,2) \simeq \mathrm{SP}_4(\mathbb{R})$. This result can be nicely linked with the following observation: the third line of the Freudenthal-Tits $2\times 2$ half split magic square
\vspace{1mm} 
\begin{center}
\begin{tabular}{ccccc}
\hline\\[-4mm]
&$\mathbb{R}$&$\mathbb{C}$&$\mathbb{H}$&$\mathbb{O}$\\[.1mm]\hline\\[-3mm]
$\mathbb{H}'$&$\mathfrak{so}(3,2)$&$\mathfrak{so}(4,2)$&$\mathfrak{so}(6,2)$&$\mathfrak{so}(10,2)$
\end{tabular}
\end{center}
\vspace{1mm}
can be reinterpreted by noting the following isomorphism
$$
\widetilde{\mathfrak{sp}}_4({\mathbb{K}})=\mathfrak{so}(n+2,2)\,
$$
with $\widetilde{\mathfrak{sp}}_4({\mathbb{K}})$ being the Sudbery symplectic algebra, where, with respect to the traditional definition, the transpose is replaced by hermitian conjugation. Recently, in \cite{MagicGroup}, it was also proposed a Lie group version of the half split $2\times 2$ magic square (see also \cite{E7,Kincaid:2014wza,Dray:2009gg} for further details and the relation with exceptional Lie algebras and groups).

Inspired by these observations we then study in details a symplectic characterization of the 4 dimensional (compactified and real) Minkowski space and superspace respectively. We argue that this approach can be also extended to 6 and possibly 10 dimensions\footnote{While the generalization to the 6 dimensional case is straightforward, further effort are needed when $d=10$ due to the notorious problem of constructing the superconformal algebra in dimension bigger than 6 \cite{nahm}; we plan to tackle this problem in a future project.}  producing thus an uniform description of ${\bf M}^{n+1,1|1}$ explicitly involving the four normed division algebras. 

More in details, while in \cite{ma2} supergrassmannians and superflags are mainly understood
as complex objects and are constructed by themselves,
in \cite{fl, flv} these objects come
together with the supergroups describing their supersymmetries, because
this is one of the main reasons of their physical significance. Furthermore, 
in \cite{fl, flv} real forms of 
both the 4d Minkowski and conformal superspaces are introduced
through suitable involutions, which are
compatible with the natural supersymmetric
action of the Poincar\'e and conformal supergroups. In the present work, we shall leave the complex structure on
the background and actually show that we can directly obtain the
real forms of the 4d Minkowski and conformal superspaces together
with their symmetry supergroups, 
without ever worrying about the 
complex field. In fact, it is very remarkable that abandoning
the complex picture, one can very quickly 
obtain the real 4d conformal and Minkowski 
 superspaces as Lagrangian supermanifold and its big cell respectively,
without the need to go to the superflag, which is undoubtedly a less
manageable geometric object, given its many defining relations (twistor
relations). 

The quantization of spacetime is an intriguing task that has been tackled from many point of views in literature. In \cite{cfln,cfl,cfl2} the authors studied the quantum deformation for the complex (chiral) Minkowski and conformal superspaces based on the general machinery developed in \cite{fquant,fquant2} for flag varieties. The more direct approach to the real Minkowski and conformal superspace we propose in this paper compares, immediately to the
one described in \cite{ma2,fl,flv} and actually we obtain the very same equations the Poincar\'e supergroup that we
find in the literature, for example in \cite{fl, flv}, but with
far less effort.  This opens the possibility to proceed further in the theory and possibly construct a quantum deformation of both real Minkowski and conformal superspaces. 
\\

The organization of the paper is as follows:\\

In Section 2 we review how finite Lorentz transformations of vectors in $n+2$ dimensional Minkowski space can be characterized by means of  "unit determinant" matrices over division algebras; this relation was originally presented in \cite{Manogue}. In Section 3 we discuss the Lie group version of the third row of $2 \times 2$ magic square \cite{MagicGroup}; in particular we show that certain type of symplectic transformations induce an $\mathrm{O}(n+2,2)$ rotation. In Section 4 we prove in details how the real form of the four dimensional Minkowski  and conformal space can be obtained as a Lagrangian manifold containing the twistor space $\mathbb{C}^4$ , 
while in Section 5 we extend this construction to the super case.

\bigskip
{\bf Acknoledgements}
We would like to thank Prof. A. Waldron, Prof. R. Bonezzi, 
Prof. J. Baez and Prof. V. S. Varadarajan for useful discussions and suggestions and especially Prof. T. Dray for its kind help in understanding the magic square of Lie groups. E.L. wishes to thank the Department of Mathematics at
the University of Bologna, for the warm ospitality during the realization of
this work. E.L. acknowledges partial support from SNF Grant No. 200020-149150/1. E.L. research was (partly) supported by the NCCR SwissMAP, funded by the Swiss National Science Foundation.

\section{Normed divison algebras and Lorentz transformations}
A normed division algebra $\mathbb{K}$ is a real algebra together with a norm $|\bullet |$ such that for all $v,w\in \mathbb{K}$ we have $|vw|=|v||w|$. By a classical result (see Hurwitz \cite{Hurwitz}) there are only four normed division algebras: the real numbers $\mathbb{R}$, the complex numbers $\mathbb{C}$, the quaternions $\mathbb{H}$, and the octonions $\mathbb{O}$, with $n:=\textrm{dim}\,\mathbb{K}=1,2,4,8$ respectively. Moreover every normed division algebra is equipped with an involutive automorphism  $v\mapsto v^*$, \emph{i.e.} the conjugation, such that $v^{**}=v$ and $(vw)^*=w^* v^*$; this leads to a natural decomposition
$$
\textrm{Re}(v):=\frac{v+v^*}{2}\,,\,\,\,\,\,\,\,\,\, \textrm{Im}(v):=\frac{v-v^*}{2}
$$
that can be used to define the inner product
$$
(v,w):=\textrm{Re}(v\,w^*)=\textrm{Re}(w\,v^*)
$$
and thus also the norm
$$
|v|:=\sqrt{v^*\,v}\,.
$$
which is real.
A division algebra element  
can be written as the linear combination $v=v^i e_{i}$ with $v^i\in\mathbb{R}$ and $i=1,..,n$. The first basis element  is the real one $e_1=1$, while the others are imaginary units $(e_i)^2=-1$, $i \neq 1$.\\
In the case $\mathbb{K}=\mathbb{H}$ the multiplication rules for the imaginary units are given by
$$
e_3e_4=-e_4e_3=e_2\,,\,e_4e_2=-e_2e_4=e_3\,,\,e_2e_3=-e_3e_2=e_4\,
$$
and similar relations, encoded in the so called Fano plane, hold also for octonions (see for example \cite{Manogue} for more details). We will then denote by $\textrm{M}_n(\mathbb{K})$ the space of $n\times n$ matrices with entries in $\mathbb{K}$ and we say that $A\in \textrm{M}_n(\mathbb{K})$ is 
\textit{hermitian} if $(A^*)^t:=A^\dag=A$; the space of $n\times n$ hermitian matrices will be denoted by $\textrm{H}_n(\mathbb{K})$. 

 \medskip
It is useful sometimes to represent any $\mathbb{H}$ valued $n\times n$ matrix as a $\mathbb{C}$ valued  $2n\times 2n$ matrix through the following map
\begin{equation}\label{map}
\begin{array}{rcrcl}
\mathrm{Z}&:& \textrm{M}_n(\mathbb{H})&\to&\textrm{M}_{2n}(\mathbb{C})\\[3mm]
&&A&\mapsto& \begin{pmatrix}  z(A)& -w^*(A)\\
w(A)& z^*(A)
 \end{pmatrix}
\end{array}
\end{equation}
where $z(A)$ and $w(A)$ mean that all the entries $v$ of the matrix $A$ are mapped into 
$$z(v)=v_1+iv_2\,,\,\,\,\,\,\,w(v)=v_3-iv_4\,$$
where $i:=e_2$ is the usual imaginary units for complex numbers.

 \medskip
Division algebras are often used in physics to easily handle supersymmetry; the minimal spinorial representations of the $n+2$ dimensional Lorentz group are, in fact, isomorphic as vector spaces to $\mathbb{K}^2$. 
 This fact relies on the observation that vectors of the $n+2$ dimensional Minkowski space ${\bf M}^{n+1,1}$ can be naturally identified with an element of $\textrm{H}_2({\mathbb{K}})$.
Consider ${\bf x}=(x_0,\cdots,x_{n+1})\in {\bf M}^{n+1,1}$ and rearrange it as follows
$$
\mathcal{X}= \begin{pmatrix}  x_0+x_{n+1}& v^* \\v&  x_0-x_{n+1}\end{pmatrix}  \in \textrm{H}_2({\mathbb{K}})
$$
where $v\in \mathbb{R},\,\mathbb{C},\, \mathbb{H},\, \mathrm{or} \,\mathbb{O}$ respectively is constructed with $x_1,\cdots,x_n$; we recognize on the top left and bottom right spot the lightcone directions we will often denote by $x_{\pm}=x_0\pm x_{n+1}$. We introduce now the so called 
\textit{trace reversal matrix} defined as follows
$$
\widetilde{\mathcal{X}}=-\begin{pmatrix}  x_0-x_{n+1}& -v^* \\-v&  x_0+x_{n+1} \end{pmatrix} \in \textrm{H}_2({\mathbb{K}})\,
$$
and  we then observe that
$$
-\textrm{det}\mathcal{X}=-x_0^2+x_{n+1}^2+|v|^2=-x_0^2+\sum_{i=1}^{n+1}x_i^2:={\bf g}({\bf x},{\bf x})
$$
or equivalently 
$$
\mathcal{X} \widetilde{\mathcal{X}}=\widetilde{\mathcal{X}}\mathcal{X}={\bf g}({\bf x},{\bf x})\,\mathbbm{1}\,
$$
with ${\bf{g}(\bullet,\bullet)}$ being the pseudo-Riemannian metric of signature $(n+1,1)$ and $\mathbbm{1}$ being the identity matrix. Note that for an $\mathbb{H}$ or $\mathbb{O}$ valued matrix we do not have a natural and well defined notion of determinant while if we restrict to the case $2\times 2$ hermitian matrices it can be unambiguously determined by the usual formula.\\
 It is instructive to reformulate the previous result introducing the symplectic matrix 
$$
\epsilon =\begin{pmatrix}  0& 1 \\-1 &  0 \end{pmatrix}
$$
so that
$$
\mathcal{X}^t\,\epsilon \,\mathcal{X} =\,\textrm{det}\mathcal{X}\,\epsilon \,.
$$

 \medskip
We consider now transformations of the form
\begin{equation}\label{transfo}
\begin{array}{rcl}
\textrm{H}_2({\mathbb{K}} )&\to& \textrm{H}_2({\mathbb{K}})\\[3mm]
\mathcal{X}&\mapsto& \lambda \mathcal{X} \lambda^{\dag}=:\mathcal{X}' \,,\,\,\,\,\,\lambda\in \textrm{M}_2(\mathbb{K})
\end{array}
\end{equation}
where extra care must be taken with octonions by requiring $(\lambda \mathcal{X}) \lambda^{\dag}=\lambda (\mathcal{X} \lambda^{\dag})$; we note that $\textrm{det}\mathcal{X}'$ is again well defined thus if we restrict the matrices $\lambda$ to those preserving the determinant under (\ref{transfo}), they will then induce a Lorentz transformation.\\
For the case of $\mathbb{R}$ or $\mathbb{C}$, this leads to the notorious statement that the special linear groups $\mathrm{SL}_2(\mathbb{R})$ and $\mathrm{SL}_2(\mathbb{C})$ double cover the 3 and 4 dimensional Lorentz group (or its connected component to the identity); one is then tempted to generalize this statement and identify the $n+2$ dimensional Lorentz transformations with unit determinant matrices over $\mathbb{K}$ but when we deal with $\mathbb{H}$ or $\mathbb{O}$ the construction of the special linear group requires further assumption and elucidation since, as we commented previously, we do not have the notion of determinant. \\
In order to get around this problem, we  note that the following relation 
$$
\textrm{det}(\lambda \mathcal{X} \lambda^{\dag} )=\textrm{det}(\lambda\lambda^\dag)\textrm{det}( \mathcal{X})
$$
is satisfied for every $\mathbb{K}$ and then the transformation (\ref{transfo}) induces a Lorentz rotation if
\begin{equation}\label{condition}
\textrm{det}(\lambda \lambda^{\dag} )=1\,;
\end{equation}
this is again unambiguous since $\lambda \lambda^{\dag}$ is hermitian.  

 \medskip
Focusing on quaternions there is an an easy way to analyze (\ref{condition}) by using the map defined in (\ref{map}); explicitly, for the case $n=2$ we get
$$
\mathrm{Z}\left[\begin{pmatrix}  a& b \\c &  d \end{pmatrix}\right] = \begin{pmatrix}  z(a)& z(b) & -w^*(a)&  -w^*(b)\\ z(c)& z(d)& -w^*(c)& -w^*(d)\\
w(a)& w(b) & z^*(a)&  z^*(b)\\ w(c)& w(d)& z^*(c)& z^*(d)
 \end{pmatrix}\,.
$$
Using this formula, it is straightforward to prove that
$$
\textrm{det}(\mathrm{Z}[\lambda])\in\mathbb{R_+}\cup\{0\}\,\,\forall \,\lambda \in\textrm{M}_2(\mathbb{H})
$$
and that
$$
\textrm{det}(\lambda \lambda^{\dag} )^2=\textrm{det}(\mathrm{Z}[\lambda \lambda^{\dag} ])=\textrm{det}(\mathrm{Z}[\lambda])\textrm{det}(\mathrm{Z}[\lambda^\dag])\,.
$$
We thus solve (\ref{condition}) by requiring $\textrm{det}(\mathrm{Z}[\lambda])= 1$ and we can then unambiguosly construct the group
$$
\mathrm{SL}(2,\mathbb{H})=\{\lambda\in \mathrm{M}_2(\mathbb{H})\,|\, \textrm{det}(\mathrm{Z}[\lambda])=1\}\,.
$$ 
When viewed in this way, $\mathrm{SL}(2,\mathbb{H})$ has a natural
real Lie group structure (Ref. \cite{vsv} Theorem 2.1.2).
It is interesting to observe that those matrices, through (\ref{transfo}), do not produce any parity or time reversal transformation, thus providing a natural double cover of the connected component to the identity of the 6 dimensional Lorentz group.

 \medskip
For the case of octonions, due to the lack of associativity, one must also impose certain compatibility conditions on the matrix $\lambda$ appearing in (\ref{transfo}); this case was originally extensively studied in \cite{Manogue} where it was also observed that not all 10 dimensional Lorentz transformations can be achieved with a single matrix  but, instead, to produce all of them one must also consider the product of $2$ octonionic valued matrices; this is the case, in fact, of transverse rotations induced by the group $\mathrm{Aut}(\mathbb{O})$; along this line, the group $\mathrm{SL}_2(\mathbb{O})$ was then explicitly characterized by a generating set of matrices.\\
In conclusion we can then establish the following isomorphism
$$\mathrm{SL}_2(\mathbb{K})\simeq \mathrm{Spin}(n+1,1)\,.$$ 

 \medskip
The matrix $\mathcal{X}$ and $\widetilde{\mathcal{X}}$ defined previosuly, have a natural interpretation in terms of Dirac gamma matrices, that we now plan to discuss. \\
Consider the spinor bundle; $S_+$ and $S_-$ are both just $\mathbb{K}^2$ as real vector spaces, but they differ as representation of $\mathrm{Spin}(n+1,1)$ (see \cite{Baez:2009xt}
 for more details). We first define how a vector acts on spinors through the gamma matrices in the Weyl base
$$
\begin{array}{rrcl}
\gamma:&{\bf M}^{n+1,1}&\to& \mathrm{Hom}(S_+,S_-)\\[2mm]
&{\bf x}\,\,\,\,\,\,\,\,\,&\mapsto &\mathcal{X}\\[3mm]
\widetilde{\gamma}:&{\bf M}^{n+1,1}&\to& \mathrm{Hom}(S_-,S_+)\\[2mm]
&{\bf x}\,\,\,\,\,\,\,\,\,&\mapsto &-\widetilde{\mathcal{X}}\,.
\end{array}
$$
In particular, our realization coincides, in 3 and 4 dimensions, with the following standard choices:
$$
\begin{array}{rcl}
\gamma&=&(\mathbbm{1},\sigma_1,\sigma_3)\\[3mm]
\widetilde{\gamma}&=&(\mathbbm{1},-\sigma_1,-\sigma_3)\end{array}
$$
and 
$$
\begin{array}{rcl}
\gamma&=&(\mathbbm{1},\sigma_1,\sigma_2,\sigma_3)\\[3mm]
\widetilde{\gamma}&=&(\mathbbm{1},-\sigma_1,-\sigma_2,-\sigma_3)
\end{array}
$$
respectively, with $\sigma_1,\sigma_2$ and $\sigma_3$ being the standard Pauli matrices. Dirac matrices can be then easily constructed as
$$
\Gamma =\begin{pmatrix}  0& \gamma \\  \widetilde{\gamma} &  0 \end{pmatrix} \in \textrm{M}_4({\mathbb{K}})\,,
$$
and in this base the chiral matrix is diagonal.
They generate the map 
$$
\Gamma:TM\to \mathrm{End}(S_-\oplus S_+)\,.
$$
explicitly given by
$$
\Gamma({\bf{x}})(\psi,\lambda)=(-\widetilde{\mathcal{X}}\lambda,\mathcal{X}\psi)
$$
with $\psi\in S_-$ and $\lambda\in S_+$.
Moreover, by construction, $\Gamma({\bf x})$ satisfies the Clifford algebra relation $\Gamma({\bf x})^2={\bf g}({\bf x},{\bf x})\mathbbm{1}$. 

\section{Symplectic realization of the conformal group}

In the following we want to extend the analysis of the previous section to the case of the conformal group $\mathrm{SO}(n+2,2)$ and relate it to a certain class of symplectic transformations. 

 \medskip
We introduce $\widetilde{{\bf M}}^{n+2,2}$ that is a pseudo-Riemannian manifold equipped with the flat metric $\mathbf{G}(\bullet,\bullet)$ with signature $(n+2,2)$, known in the conformal geometry literature as the Fefferman and Graham ambient space \cite{FG}, or better to say its flat limit. We denote by $x_{n+2}$ and $x_{0'}$ the extra space like and time like directions and we indicate by ${\bf X}=(x_{0'},x_0,x_1,\cdots,x_{n+2})$ a general vector in the ambient space; moreover we name $x_{p}=x_{0'}+ x_{n+2}$ and $x_{m}=x_{0'}- x_{n+2}$ the new pair of lightcone coordinates.

\medskip
In analogy with the case of the Lorentz group, we look for a matrix representation of the vector ${\bf X}$. To achieve this task our strategy is to use the $n+4$ dimensional Dirac gamma matrices as a guideline; we note that there is a standard method to construct them starting from the $n+2$ dimensional ones:
$$
\Upsilon=\left(\frac 1 2 \begin{pmatrix}  0&\mathbbm{1} \\  \1 &  0 \end{pmatrix} , \begin{pmatrix}  \gamma& 0 \\  0& -\widetilde{\gamma}  \end{pmatrix} ,\frac 1 2 \begin{pmatrix}  0&\1 \\  -\mathbbm{1}&  0 \end{pmatrix}  \right)
$$
and 
$$
\widetilde{\Upsilon}=\left(\frac 1 2 \begin{pmatrix}  0&\mathbbm{1} \\  \1 &  0 \end{pmatrix} , \begin{pmatrix}  \widetilde{\gamma}& 0 \\  0& -\gamma  \end{pmatrix} ,  \frac 1 2 \begin{pmatrix}  0&\1 \\  -\mathbbm{1}&  0 \end{pmatrix}  \right)\,.
$$
Inspired by this result we construct 
$$
\mathbb{X}=\begin{pmatrix}  \mathcal{X}& x_p\1 \\[2mm]  x_m \mathbbm{1}& \widetilde{\mathcal{X}}  \end{pmatrix} \in \textrm{M}_4(\mathbb{K})\,.
$$
We now look for a characterization of the metric in this representation. 
To this aim we introduce the symplectic form
$$
J=\begin{pmatrix} 0 &\mathbbm{1} \\ -\mathbbm{1} & 0 \end{pmatrix}
$$
and inspired by the  Lie algebra isomorphism 
  $$
\widetilde{\mathfrak{sp}}_4({\mathbb{K}})=\mathfrak{so}(n+2,2)
$$ with the Subdery symplectic algebra given by
$$\widetilde{\mathfrak{sp}}_4({\mathbb{K}})=\{X\in \mathrm{M}_4(\mathbb{K})\,|\, X^{\dag}J+JX=J,\,\mathrm{tr }X=0\}$$
we compute the following:
\begin{equation}\label{symplectictransfo}
\mathbb{X}^\dag J\,\mathbb{X}=\begin{pmatrix}  0& -x_mx_p\mathbbm{1}+\widetilde{\mathcal{X}}\mathcal{X} \\ x_mx_p\mathbbm{1}-\widetilde{\mathcal{X}}\mathcal{X}&  0 \end{pmatrix} =\left(\sum_{i=1}^{n+2}x_i^2-x^2_{0'}-x^2_0\right) J\,.
\end{equation}
We recognize into the bracket the ambient metric ${\bf G}({\bf X},{\bf X})$. The main observation is that the relation (\ref{symplectictransfo}) remains invariant if one transforms now $\mathbb{X} $ as follows 
$$
\mathbb{X}\to \lambda \mathbb{X} \lambda^\dag\,\,\,\,\,\textrm{with}\,\,\,\,\, \lambda\in \textrm{M}_4(\mathbb{K})\,\,\,\,\textrm{so that} \,\,\,\lambda^\dag J \lambda=J\,.
$$
  We conclude then that those types of transformations induce an $\textrm{O}(n+2,2)$ rotation and, starting from here, one can hope to construct the spinorial representation of the conformal groups $\mathrm{SO}(n+2,2)$ (or their connected component to the identity), \emph{i.e.} the spin group. For the case $\mathbb{K}=\mathbb{R}$ one obtains the notorious result $\textrm{Spin}(3,2)=\mathrm{Sp}_4(\mathbb{R})$ while when  $\mathbb{K}=\mathbb{C}$  we construct
  $$
  \widetilde{\mathrm{Sp}}_4(\mathbb{C})=\{\lambda\in \textrm{SL}_4(\mathbb{C})\,| \,\lambda^\dag J \lambda=J\}
  $$
and we use the tilde to emphasize the fact that we are using hermitian conjugation instead of the usual matrix transposition. This group double covers the connected component of the identity of $\mathrm{SO}(4,2)$ and we further analyze its properties and its Lie algebra in the next chapter.
 
  The temptation is again to extend this statement to all division algebras. In \cite{MagicGroup,E7,Kincaid:2014wza}, in fact, the authors realized $\mathrm{SO}(n+2,2)$ transformation by giving an explicit Clifford algebra description of $\mathrm{SU}(2,\mathbb{H}'\otimes \mathbb{K})$ that turns out to be equivalent to the symplectic description as $\widetilde{\mathrm{Sp}}_4( \mathbb{K})$ (see in particular \cite{E7} for a characterization of this group in the octonionic case and its connection with exceptional Lie groups); one can in conclusion establish the isomorphism 
$$
\mathrm{Spin}(n+2,2)\cong\widetilde{\mathrm{Sp}}_4( \mathbb{K})\,.$$

\medskip



\section{The $4d$ Conformal and Minkowski spaces} 
\label{cl-mink}

In
this section we would like to reinterpret the conformal space
as a Lagrangian manifold 
and the Minkowski space $\bM^{3,1}$
as a suitable big cell (hence dense) inside it. In this way the
Lagrangian manifold appears as natural compactification of the
Minkowski space $\bM^{3,1}$ and we will see that the conformal
and Poincar\'e groups will appear effortlessly in this picture as
the symmetry groups for those spaces.

\medskip
Let us recall that in the previous section, following Sudbery, we constructed the real Lie group $\tSp_4(\C)$; for convenience we report in the following its definition
$$
\tSp_4(\C)=\{\lambda \in \rSL_4(\C) \, | \,  \lambda^\dag J \lambda=J\}, \qquad\textrm{with}\qquad
J=\begin{pmatrix} 0 &\mathbbm{1} \\ -\mathbbm{1} & 0 \end{pmatrix}\,.
$$
An easy calculation shows that 
$\lambda=\begin{pmatrix} a &b \\ c & d \end{pmatrix} \in \tSp_4(\C)$ 
if and only if
$$
a^\dag c=c^\dag a, \quad b^\dag d=d^\dag b, \quad a^\dag d-c^\dag b=\mathbbm{1}
$$
where $a,b,c,d$ are $ 2 \times 2$ matrices. 

\medskip
The Lie algebra of $\tSp_4(\C)$ is explicitly given by
$$
\tsp_4(\C)=\left\{ \begin{pmatrix} x & y \\  z &  -x^\dagger  
\end{pmatrix} \in \rsl_4(\C), \,  z=z^\dagger ,\, y=y^\dagger,\, x=x^\dag
\right\}\,.
$$
In \cite{vsv} the conformal space was constructed starting from the complex conformal group $\rSL_4(\C)$ and looking at
involutions giving the real form $\rSU(2,2)$\footnote{It is not hard to prove that $\tSp_4(\C)$ and $\rSU(2,2)$ are diffeomorphic globally, and they both double cover the conformal group $\mathrm{SO}(4,2)$.  We anyhow prefer in this paper to use $\tSp_4(\C)$ that suggests a more natural generalization to the higher dimensional cases.}. In the following we instead take advantage of the symplectic interpretation of the conformal group to propose a new characterization of the 4 dimensional Minkowski space. 

 \medskip
Define now the inner product 
$$
\langle u, v \rangle:=u^{\dag}Jv\,;
$$ 
we then construct the Lagrangian $\cL$, that is,  the manifold of
totally isotropic subspaces with respect to the above inner product. Since $\mathrm{G}=\tSp_4(\C)$ acts transitively on $\cL$, we have that
\begin{equation}\label{lagreq}
\cL=\mathrm{G} \cdot \langle {\bold e}_1, {\bold e}_2 \rangle
\cong \left\{ \begin{pmatrix} a \\ c \end{pmatrix},\, a^\dag a=c^\dag a\right\}
\big/ \rGL_2(\C)\,,
\end{equation}
with $\{{\bold e}_1,{\bold e}_1,{\bold e}_1,{\bold e}_4\}$ the standard basis for $\mathbb{C}^4$. The action of $\rGL_2(\C)$ is needed to take into account 
the base change of a chosen Lagrangian subspace.
As one can readily check:
$$
\cL=\tSp_4(\C)/P$$
where 
$$
P=\left\{
\begin{pmatrix} a & b\\ 0 & (a^\dag )^{-1} \end{pmatrix} \right\}
\subset \tSp_4(\C)
$$
is the stabilizer of $\langle \eb_1, \eb_2 \rangle$.

\medskip
We define the subset $\bM^{3,1} \subset \cL$ 
consisting of those elements in $\cL$
with $a$ invertible (see eq. (\ref{lagreq})) and we write:
$$
\bM^{3,1}=\left\{ \begin{pmatrix} \mathbbm{1} \\ \mathcal{X} 
\end{pmatrix}\,\textrm{with}\,\, \mathcal{X}^\dag=\mathcal{X} \right\}\,.
$$
This expression
is obtained from (\ref{lagreq}) by right
multiplying  $\begin{pmatrix} a \\ c \end{pmatrix}$ by
 $a^{-1} \in \rGL_2(\C)$, and the interpretation of the hermitian matrix $\mathcal{X}$ is the one discussed in Section 2.
$\bM^{3,1}$ is an open dense set in $\cL$, which
is compact. $\bM^{3,1}$ is our model for the \textit{real Minkowski
space} and its compactification $\cL$ is the model for the
 \textit{real conformal space}. We now want to justify this terminology
by computing the groups that naturally act on these spaces.
We first observe that $\cL$ carries a natural action of $\tSp_4(\C)$, which we identify with the conformal group since it preserves the ambient metric, as we have shown in
Section 3. 

\medskip
We now want to compute the subgroup in $\tSp_4(\C)$
preserving $\bM^{3,1}$. We will see it becomes naturally
identified with the real Poincar\'e group and its action 
on $\bM^{3,1}$ is the correct one, restricting the action of
the conformal group $\tSp_4(\C)$ on the conformal space $\cL$.
Let $\lambda \in \tSp_4(\C)$ be such that $\lambda \cdot \bM^{3,1}=
\bM^{3,1}$, that is :
$$
\lambda \cdot {\bf x}=\begin{pmatrix} l & m \\ nl & r\end{pmatrix}
\begin{pmatrix} \mathbbm{1}  \\ \mathcal{X} 
\end{pmatrix}=\begin{pmatrix} l+m\cX \\ nl + r\cX\end{pmatrix}\,.
$$
Hence $l+m\cX$ must be invertible for all $\cX$. In particular
this gives immediately that $l$ must be invertible, so
our condition says $\mathbbm{1}+q\cX$ invertible for all $\cX$,
with $q=l^{-1}m$. By
the conditions defining $\tSp_4(\C)$, we have that $n$, $r^\dagger m$ are
hermitian and so is $r^\dagger -m^\dagger n=l^{-1}$. So $q=l^{-1}m=
r^\dagger m-m^\dagger nm$ is also hermitian. If $q \neq 0$ (i.e. $m \neq 0$), then 
$q^2>0$ so it has an eigenvalue
$k>0$. Then $\mathbbm{1}+q\cX$ is not invertible for $\cX=-k^{-1}q$.
So we conclude $m=0$. 

We have thus proven that 
the subgroup $\widehat{P}$ leaving $\bM_{3,1}$ invariant is the
the transpose of $P$ namely:
$$
\widehat{P}=\left\{
\begin{pmatrix} l & 0\\nl  & (l^\dagger )^{-1} \end{pmatrix} \right\}
\subset \tSp_4(\C)\,.
$$
It acts on $\bM^{3,1}$ as follows
$$
\begin{array}{rclcl}
\widehat{P} &\times& \bM^{3,1} & \lra & \bM^{3,1}\\ \\
  \begin{pmatrix} l & 0\\nl  & (l^\dagger )^{-1}\end{pmatrix} &,&  \cX & 
\mapsto & n + (l^{-1})^\dagger \cX l^{-1}
\end{array}
$$
and we identify the first term with space time translations, while the second contribution represents  both Lorentz rotations and dilations; we observe that this coincide with the group stabilizing a light like ray of the Dirac cone. With an abuse of terminology we call $\widehat{P}$ the 
\textit{Poincar\'e group}.


\section{The $4d$ Conformal and Minkowski superspaces}
\label{supermink}

In this section we want to generalize the results discussed in the previous 
section to the supersetting and thus construct 
the Minkowski superspace $\bM^{3,1|1}$ using the super version of $\tSp_4(\C)$,
namely $\rSpO(4|1)$, 
a real form of the symplectic-orthogonal supergroup
(for the definition of the complex $\mathrm{Osp}$ and
its equivalent  $\mathrm{SpO}$ see
\cite{ccf} Ch. 11; in \cite{fk2} it is also discussed its connection with susy curves). We shall
define this supergroup via its functor of points. 
The $R$-points 
 of the general linear supergroup consist of 
the group of invertible $m+n \times m+n$ matrices with coefficients
in the commutative superalgebra $R$ (the diagonal blocks have
even coefficients, the off diagonal blocks odd coefficients). We denote
such $R$ points with  $\rGL(m|n)(R)$.
The $R$-points of a (closed) subsupergroup $G$ of $\rGL(m|n)$ consist
of matrices in  $\rGL(m|n)(R)$ satisfying certain algebraic condition. We
are going to realize $G=\rSpO{(4|1)}$ the real symplectic-orthogonal supergroup
precisely in this way.
For all of the relevant definitions and the details we
are unable to give here,
we invite the reader to consult \cite{ccf} Ch. 1, 9, 11. 

\medskip
We then define:

$$
\rSpO{(4|1)}(R)=\{\Lambda \in \rSL_{4|1}(R) \, | \, \Lambda^\dag \mathcal{J}\Lambda=\mathcal{J}\}, \qquad\textrm{with}\qquad
\mathcal{J}=\begin{pmatrix} 0 &\mathbbm{1} & 0\\- \mathbbm{1}& 0 & 0\\ 0 & 0 & 1 \end{pmatrix}\,.
$$

\medskip
This gives us effectively a 
supergroup functor $\rSpO{(4|1)}:\salg \lra \sets$,
$R \mapsto \rSpO{(4|1)}(R)$,   
($\salg$ is the category of commutative real
superalgebras, $\sets$ is the category of sets). 
This functor is {\sl representable}, in other words, there is
a Lie supergroup corresponding to it, in the sheaf theoretic
approach (i.e. a superspace locally isomorphic to $\R^{M|N}$).
$\rSpO{(4|1)}$ is a 
closed subgroup of the complex symplectic-orthogonal
supergroup $\rSpO(4|1)$, viewed as a real supergroup. 
We shall not worry about the definition of our functors on the
arrows: such definition comes from the one of $\rGL(m|n)$ (see
\cite{ccf} Ch. 1, 11).

\medskip
If
$$
\Lambda=\begin{pmatrix} A & \alpha \\ \beta & u 
\end{pmatrix}, \quad A=\begin{pmatrix} a & b \\ c & d 
\end{pmatrix}, \quad \beta=(\beta_1, \beta_2), \quad \al=(\al_1, \al_2)^t, 
$$
with $\be_i$, $\al_i \in R^2$,
we obtain from the condition $\Lambda^\dag \mathcal{J}\Lambda=\mathcal{J}$ the following set of equations
\begin{equation}\label{spo-eq}
\left\{\begin{array}{c}A^\dagger JA+\be^\dagger \be=J \\ \\
A^\dagger J\al+\be^\dagger u=0 \\ \\
\al^\dagger JA+u^\dagger \be=0\\ \\
\al^\dagger J\al+u^\dagger u=1\end{array}
\right.
\, \iff \,
\left\{\begin{array}{c}a^\dagger c-c^\dag a +\be_1^\dagger\be_1=0 \\
a^\dagger d-c^\dagger b+\be_1^\dagger \be_2=\mathbbm{1} \\
b^\dagger c-d^\dagger a+\be_2^\dagger \be_1=-\mathbbm{1}\\
b^\dagger d-d^\dagger b+\be_2^\dagger \be_2=0\\
-c^\dagger \al_1+a^\dagger \al_2+\be_1^\dagger u=0\\
-d^\dagger \al_1+b^\dagger \al_2+\be_2^\dagger u=0\\
-\al_2^\dagger \al_1+\al_1^\dagger \al_2+u^\dagger u=1\end{array}
\right.
\end{equation}
Notice that these equations correspond for $\al=0$ and $\be=0$ to
the ones obtained in Section \ref{cl-mink}. 

\medskip
We now proceed as we did in Section \ref{cl-mink} and
look at the supermanifold $\cL$ of $2|0$ totally isotropic
subspaces. If $\{\eb_1, \eb_2, \eb_3, \eb_4, \ep\}$ 
is a basis for $\C^{4|1}$, we define
$\mathcal{L}$ as the orbit of $\langle \eb_1, \eb_2 \rangle$. This is a supermanifold
(see \cite{ccf} Proposition 9.1.4). If $R$ is a local superalgebra, we
 have
\begin{equation}\label{slag-eq}
\cL(R)=\mathrm{G} \cdot \langle \ep_1, \ep_2 \rangle=
\left\{ \begin{pmatrix} a \\ c \\ \be_1\end{pmatrix} \, \Big| \,  
a^\dagger c-c^\dagger a+\be_1^\dagger\be_1 =0 \right\}/\rGL_2(R)\,.
\end{equation}
Note that $\rGL_2(R)$ accounts as before for possible change of basis.
We then look, as in Section \ref{cl-mink}, to the open
subset of $\cL$ consisting of those subspaces corresponding
to $a$ invertible. We call it $\bM^{3,1|1}$, it will
be our model for the \textit{Minkowski superspace}, while
$\cL$ is the compactification of $\bM^{3,1|1}$ and it is the 
\textit{conformal
superspace}. By multiplying by a suitable element
of $\rGL_2(R)$ we have:
$$
\bM^{3,1|1}(R)=\left\{ \begin{pmatrix} \mathbbm{1} \\ \mathcal{Y} \\ \zeta \end{pmatrix} \, \Big|\, 
\mathcal{Y}^\dagger =\mathcal{Y}+\zeta^\dagger\zeta \right\}\,.
$$
Here $R$ is a commutative superalgebra, not necessarily local as before.
Notice that $\mathcal{Y}=ca^{-1}$, $\zeta=\be_1a^{-1}$ with respect to
the expression in (\ref{slag-eq}). Hence the equation is obtained
immediately from (\ref{slag-eq}) by setting $a=1$. This is precisely the condition found in
\cite{flv}. Notice that here the condition is coming naturally
from the context we have chosen, while in \cite{flv}
the same condition is obtained with more effort, through
an involution of the conformal superspace. 
With this approach we are able to
compute directly the real Minkowski superspace without resorting to
the superflag.

\medskip
We now turn and examine the Poincar\'e supergroup. We want
a supergroup acting on $\bM^{3,1|1}$. 
We notice that the supergroup functor
$$
\widehat{sP}(R)=\left\{ \begin{pmatrix} L & 0 & 0 \\ M & R & R\phi \\ 
d\chi & 0 & d \end{pmatrix} \right\}
$$
leaves $\bM^{3,1|1}$ invariant, it is representable (it is a closed
subsupergroup of $\tSp(4|1)$ and its reduced group is the Poincar\'e group.
(We use the notation as in \cite{flv} so to make the
comparison easier). 
We take then $\widehat{sP}(R)$ as our definition for 
the Poincar\'e supergroup. 
Applying the equations in (\ref{spo-eq}) to $\widehat{sP}(R)$ we obtain:
$$
R=(L^\dagger )^{-1}, \quad \phi=\chi^\dagger, \quad
ML^{-1}=(ML^{-1})^\dagger + (L^\dagger)^{-1}\chi^\dagger \chi L^{-1} \,.
$$
Hence 
$$
\widehat{sP}(R)=\left\{ \begin{pmatrix} L & 0 & 0 \\ M & (L^\dagger )^{-1} & 
(L^\dagger )^{-1}\chi^\dagger  \\ 
d\chi & 0 & d \end{pmatrix} \right\}
$$
which is precisely the Poincare' supergroup as given in \cite{flv}.

\medskip
The action on $\bM^{3,1|1}$ can then be readily computed, and it yields:
$$
\begin{array}{rclcl}
\widehat{sP} &\times& \bM^{3,1|1} & \lra & \bM^{3,1|1}\\ \\
  \scalebox{.8}{$\begin{pmatrix} L & 0 & 0 \\ M & (L^\dagger )^{-1} & 
(\chi L^{-1})^\dag\  \\ 
d\chi & 0 & d \end{pmatrix}$} &,&   \scalebox{.8}{$\begin{pmatrix} \mathbbm{1}\\[2mm]  \mathcal{Y} \\[2mm] \zeta \end{pmatrix} $}& 
\mapsto &  \scalebox{.8}{$\begin{pmatrix}  \mathbbm{1}\\[2mm] ML^{-1}+(L^\dag)^{-1}\mathcal{Y}L^{-1}  +(\chi L^{-1})^\dag\zeta L^{-1}\\[2mm] d\chi L^{-1}+d\zeta L^{-1} \end{pmatrix} $}
\end{array}
$$
as expected.

\end{document}